\documentclass[journal]{IEEEtran}

\usepackage{graphicx,latexsym,color,wrapfig}
\usepackage{amsmath,amsthm,amsfonts}
\usepackage{latexsym}
\usepackage{cite,citesort}

\usepackage{bm}

\newcommand{\ie}{\textit{i.e.}\/, }
\newcommand{\eg}{\textit{e.g.}\/, }
\newcommand{\cf}{\textit{cf.}\/, }

\providecommand*{\mrm}[1]{\mathrm{#1}}
\providecommand*{\unit}[1]{\ensuremath{\mrm{\,#1}}}
\providecommand*{\eu}{\ensuremath{\mrm{e}}}
\providecommand*{\iu}{\ensuremath{\mrm{i}}}

\renewcommand{\Re}{\operatorname{Re}}	
\renewcommand{\Im}{\operatorname{Im}}	

\newtheorem{proposition}{\em Proposition}[section]

\begin{document}

\title{Fisher Information for Inverse Problems and \\ Trace Class Operators}

\author{Sven~Nordebo$^{*}$, \and Mats~Gustafsson$^{\dagger}$, \and Andrei~Khrennikov$^{*}$, \and B\"{o}rje~Nilsson$^{*}$, \and Joachim~Toft$^{*}$
\thanks{$^{*}$Sven Nordebo, Andrei Khrennikov, B\"{o}rje Nilsson and Joachim Toft are with the 
School of Computer Science, Physics and Mathematics, Linn\ae us University, 351 95 V\"{a}xj\"{o}, Sweden.
E-mail: \{sven.nordebo,andrei.khrennikov,borje.nilsson,joachim.toft\}@lnu.se.}
\thanks{$^{\dagger}$Mats Gustafsson is with the 
Department of Electrical and Information Technology, Lund University, Box 118, SE-221 00 Lund, Sweden. E-mail: mats.gustafsson@eit.lth.se.}
}


\date{\today}

\maketitle 

\begin{abstract}
This paper provides a mathematical framework for Fisher information analysis for inverse problems 
based on Gaussian noise on infinite-dimensional Hilbert space. The covariance operator for the
Gaussian noise is assumed to be trace class, and the Jacobian of the forward operator Hilbert-Schmidt.
We show that the appropriate space for defining the Fisher information is given by the Cameron-Martin space.
This is mainly because the range space of the covariance operator always is strictly smaller than the Hilbert space.
For the Fisher information to be well-defined, it is furthermore required that the range space of the Jacobian is contained in the Cameron-Martin space.
In order for this condition to hold and for the Fisher information to be trace class, a sufficient condition is formulated
based on the singular values of the Jacobian as well as of the eigenvalues of the covariance operator, together with some regularity assumptions 
regarding their relative rate of convergence. 
An explicit example is given regarding an electromagnetic inverse source problem with ``external'' spherically isotropic noise,
as well as ``internal'' additive uncorrelated noise.
\end{abstract}


\begin{IEEEkeywords}
Inverse problems, Fisher information, Cram\'{e}r-Rao lower bound, trace class operators.
\end{IEEEkeywords}

\section{Introduction}
An important issue within inverse problems and imaging \cite{Kirsch1996} is to develop an appropriate treatment of the uncertainty associated 
with the related quantity of interest (the state), and the observation process, respectively.  
Statistical methods provide many useful concepts and tools in this regard such as identifiability, sufficient statistics, 
Fisher information, statistically based decision rules and Bayesian estimation, see \eg \cite{Evans+Stark2002,Kaipio+Somersalo2005,Tarantola2005}. 
The Bayesian view is \eg based on the assumption that uncertainty always can be represented as a probability distribution.
Hence, statistical information about the state can be retrieved by means of Bayesian filtering such as Kalman filters, extended Kalman filters 
and particle filters, see e.g. \cite{Kaipio+Somersalo2005,Tarantola2005}.

On the other hand, the conditional Fisher information can be a useful ``deterministic'' analysis tool in various inverse problem applications
such as with material measurements \cite{Sjoberg+Larsson2011}, parametric shape estimation \cite{Ye+etal2003},
sensitivity analysis and preconditioning in microwave tomography \cite{Nordebo+etal2008,Nordebo+etal2010}, 
and in electrical impedance tomography \cite{Nordebo+etal2010b}. 
The Cram\'{e}r-Rao lower bound can furthermore be used to quantify the trade-off between the accuracy and the resolution of a linear or non-linear inverse problem 
\cite{Nordebo+etal2008,Nordebo+etal2010,Nordebo+etal2010b,Nordebo+Gustafsson2006,Gustafsson+Nordebo2006a}, 
and it complements the deterministic upper bounds and the L-curve techniques that are employed with
linearized inversion, see \eg \cite{Kirsch1996,Hansen2010}. 
The Fisher information is also a tool that can be exploited for optimal experimental design, see \eg \cite{Pronzato2008}.

The Fisher information analysis has mostly been given in a finite-dimensional setting, see \eg \cite{Kay1993,Scharf1991}. 
One of the very few exceptions can be found in \eg \cite{VanTrees1968} where the Fisher information integral operator for 
an infinite-dimensional parameter function has been employed in the estimation of wave forms (or random processes). 
An infinite-dimensional Fisher information operator has previously been exploited in the context of inverse problems in \eg
\cite{Nordebo+etal2010,Nordebo+etal2010b,Nordebo+etal2011c}, and the connection to the Singular Value Decomposition (SVD) has 
been established. However, so far, a rigorous treatment of Gaussian noise on infinite-dimensional Hilbert spaces has been lacking.

A Fisher information analysis for inverse problems based on Gaussian noise on infinite-dimensional Hilbert space
\cite{Hairer2009,Bogachev1998,Khrennikov2007,Gikhman+Skorokhod2004,Miller1974} requires a careful study
of the range spaces and the spectrum of the operators related to the covariance of the noise, and the Jacobian of the forward operator, 
respectively. It is assumed here that the covariance operator is trace class and the Jacobian is Hilbert-Schmidt \cite{Reed+Simon1980}.
Since the range space of the covariance operator always is strictly smaller than the Hilbert space,
the appropriate space for defining the Fisher information operator is given by the Cameron-Martin space \cite{Hairer2009,Bogachev1998}.
In order for the Fisher information to be well-defined, it is furthermore required that the range space of the Jacobian is contained
in the Cameron-Martin space. In order for this condition to hold and for the Fisher information to be trace class, it is shown in this paper that
it is sufficient that the singular values of the Jacobian and the eigenvalues of the covariance operator satisfy some regularity assumptions 
regarding their relative rate of convergence. 
It is also shown how the Cram\'{e}r-Rao lower bound for parameter estimation based on finite-dimensional eigenspaces is closely
linked to the Singular Value Decomposition (SVD), and the truncated pseudo-inverse based on the Cameron-Martin space.

The situation becomes particularly simple if the covariance operator and the Jacobian share the same left-singular vectors, 
implying that a sequence of finite-dimensional subspace estimators (truncated pseudo-inverses) may readily be analyzed. 
In this case, the eigenvalues of the Fisher information is simply the ratio between the squared singular values of the Jacobian,
and the eigenvalues of the covariance operator. Two important special cases arises: 
1) The infinite-dimensional Fisher information operator exists and is trace class, and
the corresponding pseudo-inverse (and the Cram\'{e}r-Rao lower bound) exists only for finite-dimensional subspaces.
2) The infinite-dimensional pseudo-inverse (and the Cram\'{e}r-Rao lower bound) exists, and the corresponding
Fisher information operator exists only for finite-dimensional subspaces.
A concrete example is considered regarding an electromagnetic inverse source problem with ``external'' spherically isotropic noise,
as well as ``internal'' uncorrelated noise.
The example illustrates how the singular values of the Jacobian, the eigenvalues of the covariance operator
and the internal noise variance can be analyzed and interpreted in terms of the Cram\'{e}r-Rao lower bound.

\section{Cram\'{e}r-Rao Lower Bound for Inverse Problems}\label{sect:CRB}
\subsection{The forward model}
Below, two common situations will be considered in parallel, where the parameter space (describing the quantity of interest) is real or complex, respectively.
The observation process (the measurement) is assumed to be executed in the frequency-domain with complex valued data.
Hence, an inverse problem is considered below where $\xi$ denotes the complex valued measurement and 
$\theta$ the real or complex valued parameter function to be estimated.
The forward operator is defined as the mapping
\begin{equation}
\psi(\theta):\Theta\rightarrow {\cal H},
\end{equation}
where $\Theta$ and ${\cal H}$ are separable Hilbert spaces \cite{Kreyszig1978}.
The measurement space ${\cal H}$ is complex, and the parameter space $\Theta$ is either real or complex.
The same notation $\langle \cdot,\cdot \rangle$ is employed here to denote the scalar products in both spaces. 

Let $\Re\{\cdot\}$ denote the real part of $\{\cdot\}$.
Note that $\Re\langle \xi_1,\xi_2 \rangle$ is a scalar product defined on the elements  $\xi_1$ and $\xi_2$ of ${\cal H}$ interpreted as a Hilbert space over the 
real scalar field. In this space, the vectors $\xi$ and $\iu\xi$ are orthogonal, \ie $\Re \langle \xi,\iu\xi \rangle=\Re \iu\langle \xi,\xi \rangle=0$,
and the norm is preserved, \ie $\Re \langle \xi,\xi \rangle=\langle \xi,\xi \rangle=\|\xi\|^2$.
Hence, the Hilbert space ${\cal H}$ over the complex scalar field with basis vectors $\{e_i\}$ is isomorphic
with the Hilbert space ${\cal H}$ over the real scalar field, and with basis vectors given by $\{e_i\}$ and $\{\iu e_i\}$.

It is assumed that the forward operator $\psi(\theta)$ is Fr\'{e}chet differentiable \cite{Kirsch1996} in a neighborhood of the
(background) parameter vector $\theta_{\rm b}\in\Theta$.
The first order variation $\delta\!\psi$ can then be represented as 
\begin{equation}\label{eq:Jdef1}
\delta\!\psi={\cal J} \delta\!\theta,
\end{equation}
where the bounded linear operator ${\cal J}:\Theta\rightarrow{\cal H}$ 
is the Fr\'{e}chet derivative (or Jacobian) of the operator $\psi(\theta)$ evaluated at $\theta_{\rm b}$, and
$\delta\! \theta\in \Theta$ is an incremental parameter function where $\theta=\theta_{\rm b}+\delta\!\theta$.
Let ${\cal J}^*$ denote the Hilbert adjoint operator to ${\cal J}$.
It is assumed that the Jacobian ${\cal J}$ is Hilbert-Schmidt \cite{Reed+Simon1980}, so that $\textrm{tr}\{{\cal J}^*{\cal J}\}<\infty$ and 
both ${\cal J}$ and ${\cal J}^*$ are compact.

For a bounded linear operator, the Hilbert-adjoint operator exists and is unique provided that the Hilbert spaces are defined
over the same (real or complex) scalar field \cite{Kreyszig1978}. Hence, in the case when $\Theta$ is real
the Hilbert-adjoint operator ${\cal J}^*$ is defined by the property
\begin{equation}\label{eq:Jadjdef1}
\Re\langle {\cal J}\delta\!\theta,\xi \rangle=\langle \delta\!\theta,{\cal J}^*\xi\rangle,
\end{equation}
and $\langle {\cal J}\delta\!\theta,\xi \rangle=\langle \delta\!\theta,{\cal J}^*\xi\rangle$ when $\Theta$ is complex.
As \eg if the Jacobian ${\cal J}$ can be represented by a (finite or infinite-dimensional) matrix ${\bf J}$, 
the adjoint ${\cal J}^*\xi$ is given by $\Re\{{\bf J}^{\rm H}\bm{\xi}\}$ and ${\bf J}^{\rm H}\bm{\xi}$
with respect to the real and complex spaces $\Theta$, respectively, where $\bm{\xi}$ denotes the representation of $\xi\in{\cal H}$, and
$(\cdot)^{\rm H}$ the Hermitian transpose.

The Singular Value Decomposition (SVD) of the Jacobian ${\cal J}$ can be expressed as follows.
Assume first that $\Theta$ is complex.
Let the singular system for the linear compact operator ${\cal J}$ be given by
\begin{equation}\label{eq:singsystem}
\left\{\begin{array}{l}
{\cal J}v_i=\sigma_i u_i, \vspace{0.2cm}\\
{\cal J}^*u_i=\sigma_i v_i,
\end{array}\right.
\end{equation}
where $\{u_i\}$ and $\{v_i\}$ are orthonormal systems and $\sigma_i$ are the singular values 
arranged in decreasing order, and where $\sigma_i\rightarrow 0$ as $i\rightarrow\infty$ if the number of non-zero singular values
is infinite \cite{Kirsch1996}.
The operator ${\cal J}$ can then be represented as
\begin{equation}\label{eq:Jdef}
{\cal J}\delta\!\theta=\sum_{i\in I}\sigma_iu_i\langle v_i,\delta\!\theta \rangle,
\end{equation}
the adjoint operator ${\cal J}^*$ as
\begin{equation}\label{eq:Jadjdef}
{\cal J}^*\xi=\sum_{i\in I}\sigma_iv_i\langle u_i,\xi \rangle,
\end{equation}
and the pseudo-inverse ${\cal J}^{+}$  as
\begin{equation}\label{eq:Jpdef}
{\cal J}^{+}\xi=\sum_{i\in I}\frac{1}{\sigma_i}v_i\langle u_i,\xi \rangle,
\end{equation}
where $I$ is the index set (finite or infinite) corresponding to the non-zero singular values. Here, $\langle u_i,\xi \rangle$ is replaced for
$\Re\langle u_i,\xi \rangle$ when $\Theta$ is real.
Note that the expression (\ref{eq:Jpdef}) is valid
provided that the pseudo-inverse exists, and the series $\sum_i|\langle u_i,\xi \rangle|^2/\sigma_i^2$ converges \cite{Kirsch1996}. 
A truncation of (\ref{eq:Jpdef}) using a finite number $r$ of non-zero singular values is referred to as a truncated pseudo-inverse.
The corresponding regularization strategy is denoted
\begin{equation}\label{eq:Jpdefr}
{\cal J}^{+}_{r}\xi=\sum_{i=1}^{r}\frac{1}{\sigma_i}v_i\langle u_i,\xi \rangle,
\end{equation}
where $r$ is the regularization parameter.
It is readily seen that 
\begin{equation}\label{eq:Jpdefrinv}
\lim_{r\rightarrow\infty}{\cal J}^{+}_{r}{\cal J}\delta\!\theta=\overline{\delta\!\theta},
\end{equation}
where $\overline{\delta\!\theta}$ belongs to the orthogonal complement of the null space of the operator ${\cal J}$,
\ie $\overline{\delta\!\theta}\in{\cal N}({\cal J})^{\perp}$, see \eg \cite{Kirsch1996}.
Note that the SVD gives an orthogonal decomposition of the two Hilbert spaces as follows
\begin{equation}
\left\{\begin{array}{l}
\Theta=\overline{\textrm{span}\{v_i\}_{i\in I}}\oplus {\cal N}\{{\cal J}\}, \vspace{0.2cm}\\
{\cal H}=\overline{\textrm{span}\{u_i\}_{i\in I}}\oplus {\cal N}\{{\cal J^*}\},
\end{array}\right.
\end{equation}
where $\overline{\textrm{span}\{\cdot\}}$ denotes the closure of the linear span of the vectors in $\{\cdot\}$ and $\oplus$ the direct sum \cite{Kirsch1996}.

\subsection{Fisher information for the Cameron-Martin space}
Consider the statistical measurement model
\begin{equation}\label{eq:statmodel}
\xi=\psi(\theta)+w,
\end{equation}
where $\xi$ is the measured response, $\psi(\theta)$ the forward model
and $w\in{\cal H}$ is zero mean complex Gaussian noise.
Here, $w$ is an observation of a complex Gaussian stochastic process defined on the Hilbert space ${\cal H}$ 
\cite{Hairer2009,Bogachev1998,Khrennikov2007,Gikhman+Skorokhod2004,Miller1974}.
Let ${\cal E}$ denote the expectation operator with respect to the complex Gaussian measure $\mu$
and $B$ the corresponding covariance operator. It is assumed that the positive operator $B$ is trace class
so that ${\cal E}\{\|w\|^2\}=\textrm{tr}\{B\}$ for any $w\in{\cal H}$, and hence that $B$ is self-adjoint and compact \cite{Reed+Simon1980}.
It follows from the Hilbert-Schmidt theorem \cite{Reed+Simon1980} that there
exists a complete orthonormal basis $\{\phi_j\}$ for ${\cal H}$, such that
\begin{equation}\label{eq:Beig}
B\phi_j=\lambda_j\phi_j,
\end{equation}
where $\phi_j$ are the eigenvectors of $B$ and $\lambda_j$ the non-negative eigenvalues, and where $\lambda_j\rightarrow 0$ as $j\rightarrow\infty$
if the number of non-zero eigenvalues is infinite.
Let $J$ denote the index set (finite or infinite) corresponding to the eigenvectors with non-zero eigenvalues $\lambda_j>0$.
The Hilbert space ${\cal H}$ is then given by the following orthogonal decomposition
\begin{equation}
{\cal H}=\overline{\textrm{span}\{\phi_j\}_{j\in J}}\oplus {\cal N}\{B\}.
\end{equation}

The linear functionals $\langle \xi,w\rangle$ and $\langle\eta,w\rangle$ are complex Gaussian stochastic variables with zero mean,
and with the covariance properties
\begin{equation}\label{eq:covprop}
\left\{\begin{array}{l}
{\cal E}\{\langle \xi,w\rangle\langle \eta,w\rangle^* \}=\langle \xi,B\eta \rangle, \vspace{0.2cm}\\
{\cal E}\{\langle \xi,w\rangle\langle \eta,w\rangle \}=0,
\end{array}\right.
\end{equation}
for any $\xi,\eta\in{\cal H}$, and where $(\cdot)^*$ denotes the complex conjugate \cite{Hairer2009,Bogachev1998,Khrennikov2007,Gikhman+Skorokhod2004,Miller1974}.

The description of the stochastic process on ${\cal H}$ is now complete. However, since the range of the operator $B$ always is strictly
smaller than ${\cal H}$ for infinite-dimensional spaces\footnote{To see this, assume that ${\cal R}\{B\}={\cal H}$. Now, $B$ is trace class and thus bounded,
and $B^{-1}$ exists since $B>0$. It then follows by the bounded inverse theorem \cite{Kreyszig1978}, that $B^{-1}$ is bounded. 
Since the composition of a trace class operator with a bounded operator is trace class \cite{Reed+Simon1980}, the equality $BB^{-1}=I$ gives a 
contradiction since the identity operator is not trace class for infinite-dimensional spaces.}, 
it is not possible to define a non-singular Maximum Likelihood (ML) criterion\cite{Kay1993,Scharf1991,VanTrees1968}, 
or trace class Fisher information based on the whole of ${\cal H}$.
Moreover, in order to be able to define the Fisher information in trace class, the range of the Jacobian ${\cal J}$ also needs to
be restricted to this smaller subspace. 
The proper subspace for this purpose is the Cameron-Martin space ${\cal H}_{\mu}$ \cite{Hairer2009,Bogachev1998}, which can be constructed as follows.
Consider the covariance operator $B$ restricted to the range of $B$, \ie
\begin{equation}
B:{\cal R}\{B\}\rightarrow {\cal R}\{B\},
\end{equation}
then the (self-adjoint) inverse $B^{-1}$ exists on this subspace since $B>0$, and hence $B^{-1}>0$ on ${\cal R}\{B\}$.

Define the scalar product
\begin{equation}\label{eq:scalardef1}
\langle \xi,\eta \rangle_{\mu}=\Re \langle \xi,B^{-1}\eta \rangle,
\end{equation}
if $\Theta$ is real, and $\langle \xi,\eta \rangle_{\mu}=\langle \xi,B^{-1}\eta \rangle$ if $\Theta$ is complex,
and where $\xi,\eta\in{\cal R}\{B\}$. The Cameron-Martin space ${\cal H}_{\mu}$ is the completion of ${\cal R}\{B\}$ with respect to 
the norm induced by the scalar product $\langle \xi,\eta \rangle_{\mu}$, see \cite{Hairer2009,Bogachev1998}.
It can be shown \cite{Hairer2009,Bogachev1998} that ${\cal H}_{\mu}$ is also given by
\begin{equation}
\left\{\begin{array}{l}
\displaystyle {\cal H}_{\mu}=\{\xi\in{\cal H}| \sum_{j\in J}\frac{1}{\lambda_j}|\langle \phi_j,\xi \rangle|^2<\infty\}, \vspace{0.2cm} \\
{\cal H}_{\mu}={\cal R}\{B^{1/2}\},
\end{array}\right.
\end{equation}
where ${\cal R}\{B\}\subset {\cal H}_{\mu}\subset {\cal H}$.

It is assumed that the range of the Jacobian ${\cal J}$ is restricted as ${\cal R}\{{\cal J}\}\subset {\cal R}\{B\}$, 
implying that ${\cal R}\{{\cal J}\}\subset {\cal H}_{\mu}$
\footnote{The main results of this paper can be generalized to larger class of Jacobians ${\cal J}$ having the image in $H_\mu$ (and not just in its proper
subspace ${\cal R}\{B\}$. However, this will make the paper more complicated mathematically, 
since one cannot proceed with continuous linear functionals given by the scalar product, and
measurable linear functionals (square integrable with respect to the Gaussian measure $\mu$) have to be used.}.
Hence, it is sufficient to consider the mapping
\begin{equation}\label{eq:suffmap}
{\cal J}:\Theta\rightarrow {\cal H}_{\mu}.
\end{equation}
Since ${\cal H}_{\mu}$ is a Hilbert space on its own right and ${\cal J}$ a bounded linear operator, the Hilbert adjoint operator ${\cal J}_{\mu}^*$ exists
and is defined by the property 
\begin{equation}
\langle {\cal J}\delta\!\theta,\xi \rangle_{\mu}=\langle \delta\!\theta,{\cal J}_{\mu}^*\xi \rangle,
\end{equation}
where $\delta\!\theta\in\Theta$ and $\xi\in{\cal H}_{\mu}$.
It is readily seen that ${\cal J}_{\mu}^*={\cal J}^*B^{-1}$ on ${\cal R}\{B\}$, where ${\cal J}^*$ was defined as in (\ref{eq:Jadjdef1}).

The Fisher information operator ${\cal I}:\Theta\rightarrow\Theta$ can now be defined as
\begin{equation}\label{eq:Fishdef1}
{\cal I}=2{\cal J}_{\mu}^*{\cal J}=2{\cal J}^*B^{-1}{\cal J},
\end{equation}
when $\Theta$ is real, and ${\cal I}={\cal J}_{\mu}^*{\cal J}$ when $\Theta$ is complex, see also \cite{VanTrees1968,Kay1993,Nordebo+etal2010b,Nordebo+etal2010}.
It is assumed that the operator ${\cal J}$ is Hilbert-Schmidt with respect to the Cameron-Martin space ${\cal H}_{\mu}$,
and hence that ${\cal J}$ is compact with respect to the norm induced by the scalar product $\langle \xi,\eta \rangle_{\mu}$.
The Fisher information operator is thus assumed to be trace class.
Let $\tilde\mu_i$ and $\tilde v_i$ denote the eigenvalues and the eigenvectors of the Fisher information operator ${\cal I}$, respectively, \ie
\begin{equation}\label{eq:Fishdef2}
{\cal I}\tilde v_i=\tilde\mu_i\tilde v_i,
\end{equation}
where the eigenvectors $\tilde v_i$ are assumed to be orthonormal, \ie $\langle \tilde v_i,\tilde v_j\rangle=\delta_{ij}$.

The Singular Value Decomposition (SVD) of the Jacobian ${\cal J}$ with respect to the Cameron-Martin space ${\cal H}_{\mu}$, can now
be expressed as follows.
The singular system for the linear compact operator ${\cal J}$ is given by
\begin{equation}\label{eq:singsystemmu}
\left\{\begin{array}{l}
{\cal J}\tilde v_i=\tilde \sigma_i \tilde u_i, \vspace{0.2cm}\\
{\cal J}_{\mu}^*\tilde u_i=\tilde\sigma_i \tilde v_i,
\end{array}\right.
\end{equation}
where $\{\tilde u_i\}$ and $\{\tilde v_i\}$ are orthonormal systems and $\tilde\sigma_i$ are the singular values 
arranged in decreasing order, and where $\tilde\sigma_i\rightarrow 0$ as $i\rightarrow\infty$
if the number of non-zero singular values is infinite \cite{Kirsch1996}.
Note that the eigenvalues of the Fisher information operator are given by $\tilde\mu_i=2\tilde\sigma_i^2$ when $\Theta$ is real,
and $\tilde\mu_i=\tilde\sigma_i^2$ when $\Theta$ is complex.

The operator ${\cal J}$ can be represented as
\begin{equation}\label{eq:Jdefmu}
{\cal J}\delta\!\theta=\sum_{i\in \tilde I}\tilde\sigma_i \tilde u_i\langle \tilde v_i,\delta\!\theta \rangle,
\end{equation}
the adjoint operator ${\cal J}_{\mu}^*$ as
\begin{equation}\label{eq:Jadjdefmu}
{\cal J}_{\mu}^*\xi=\sum_{i\in \tilde I}\tilde \sigma_i\tilde v_i\langle \tilde u_i,\xi \rangle_{\mu},
\end{equation}
and the pseudo-inverse ${\cal J}_{\mu}^{+}$  as
\begin{equation}\label{eq:Jpdefmu}
{\cal J}_{\mu}^{+}\xi=\sum_{i\in \tilde I}\frac{1}{\tilde\sigma_i}\tilde v_i\langle \tilde u_i,\xi \rangle_{\mu},
\end{equation}
where $\tilde I$ is the index set (finite or infinite) corresponding to the non-zero singular values. 
The expression (\ref{eq:Jpdefmu}) is valid
provided that the pseudo-inverse exists and the series $\sum_i|\langle \tilde u_i,\xi \rangle_{\mu}|^2/\tilde\sigma_i^2$ converges \cite{Kirsch1996}. 
A truncation of (\ref{eq:Jpdefmu}) using a finite number $r$ of non-zero singular values is referred to as a truncated pseudo-inverse.
The corresponding regularization strategy is denoted
\begin{equation}\label{eq:Jpdefrmu}
{\cal J}_{\mu r}^{+}\xi=\sum_{i=1}^{r}\frac{1}{\tilde\sigma_i}\tilde v_i\langle \tilde u_i,\xi \rangle_{\mu},
\end{equation}
where $r$ is the regularization parameter.
It is readily seen that 
\begin{equation}\label{eq:Jpdefrinvmu}
\lim_{r\rightarrow\infty}{\cal J}_{\mu r}^{+}{\cal J}\delta\!\theta=\overline{\delta\!\theta},
\end{equation}
where $\overline{\delta\!\theta}\in{\cal N}({\cal J})^{\perp}$, see \eg \cite{Kirsch1996}.
Note also that the SVD gives an orthogonal decomposition of the two Hilbert spaces as follows
\begin{equation}
\left\{\begin{array}{l}
\Theta=\overline{\textrm{span}\{\tilde v_i\}_{i\in\tilde I}}\oplus {\cal N}\{{\cal J}\}, \vspace{0.2cm}\\
{\cal H}_{\mu}=\overline{\textrm{span}\{\tilde u_i\}_{i\in\tilde I}}\oplus {\cal N}\{{\cal J_{\mu}^*}\}.
\end{array}\right.
\end{equation}

\subsection{The Cram\'{e}r-Rao lower bound based on trace class operators}

Consider the statistical measurement model given by (\ref{eq:statmodel}).
Consider also the SVD of the Jacobian ${\cal J}$ with respect to the Cameron-Martin space ${\cal H}_{\mu}$ 
given in (\ref{eq:singsystemmu}), where ${\cal J}$ is evaluated at the fixed background parameter vector $\theta_{\rm b}$, 
and let $\theta=\theta_{\rm b}+\delta\!\theta$.
Any incremental parameter vector $\delta\!\theta\in\Theta$ can be uniquely decomposed as
\begin{equation}
\delta\!\theta=\overline{\delta\!\theta}+\theta_0=\sum_{i\in \tilde I}\vartheta_i \tilde v_i + \theta_0,
\end{equation}
where $\overline{\delta\!\theta}\in{\cal N}({\cal J})^{\perp}$ and $\theta_0\in{\cal N}({\cal J})$.
The {\em principal parameters} of interest are given by
\begin{equation}\label{eq:principaldef}
\vartheta_i=\langle \tilde v_i,\delta\!\theta \rangle=\langle \tilde v_i,\overline{\delta\!\theta} \rangle,
\end{equation}
where $i\in \tilde I$.

Assume first that $\Theta$ is real.
Consider the discrete (countable) measurement model that is obtained by applying the scalar products $\Re\langle B^{-1}\tilde u_i,\cdot\rangle$ 
to the statistical measurement model (\ref{eq:statmodel}). Note that $\tilde u_i\in {\cal R}\{{\cal J}\}\subset{\cal R}\{B\}$ so that
$B^{-1}\tilde u_i$ is uniquely defined.
The discrete measurement model is hence given by
\begin{equation}\label{eq:suffstat1}
\Re\langle B^{-1}\tilde u_i,\xi \rangle=\Re\langle B^{-1}\tilde u_i,\psi(\theta) \rangle+\Re\langle B^{-1}\tilde u_i,w \rangle,
\end{equation}
where $i\in \tilde I$. The covariance of the noise term is given by
\begin{equation}\label{eq:covariance1}
\begin{array}{l}
{\cal E}\{\Re\langle B^{-1}\tilde u_i,w \rangle \Re\langle B^{-1}\tilde u_j,w \rangle\} \vspace{0.2cm} \\
\hspace{1cm} =\displaystyle\frac{1}{2}\Re\langle \tilde u_i,B^{-1}\tilde u_j \rangle
=\frac{1}{2}\langle \tilde u_i,\tilde u_j \rangle_{\mu}=\frac{1}{2}\delta_{ij},
\end{array}
\end{equation}
where $i,j\in \tilde I$ and where (\ref{eq:covprop}) and (\ref{eq:scalardef1}) have been used.
When $\Theta$ is complex, the corresponding discrete measurement model is given by
\begin{equation}\label{eq:suffstat2}
\langle B^{-1}\tilde u_i,\xi \rangle=\langle B^{-1}\tilde u_i,\psi(\theta) \rangle+\langle B^{-1}\tilde u_i,w \rangle,
\end{equation}
where $i\in \tilde I$. The covariance of the noise term is given by
\begin{equation}\label{eq:covariance2}
\begin{array}{l}
{\cal E}\{\langle B^{-1}\tilde u_i,w \rangle \langle B^{-1}\tilde u_j,w \rangle^*\} \vspace{0.2cm} \\
\hspace{1cm} 
=\langle \tilde u_i,B^{-1}\tilde u_j \rangle=\langle \tilde u_i,\tilde u_j \rangle_{\mu}=\delta_{ij},
\end{array}
\end{equation}
where $i,j\in \tilde I$. 

Assume once again that $\Theta$ is real.
By using $\delta\!\theta=\tilde v_j\textrm{d}\vartheta_j$ and letting $\textrm{d}\vartheta_j\rightarrow 0$,
it is readily seen that
\begin{equation}\label{eq:difforthorel}
\begin{array}{l}
\displaystyle\frac{\partial}{\partial\vartheta_j}\Re\langle B^{-1}\tilde u_i,\psi(\theta) \rangle\vspace{0.2cm} \\
\hspace{1cm}=\Re\langle B^{-1}\tilde u_i, {\cal J}\tilde v_j\rangle=\Re \langle B^{-1}\tilde u_i, \tilde\sigma_j \tilde u_j\rangle=\tilde\sigma_j\delta_{ij},
\end{array}
\end{equation}
where $i,j\in \tilde I$. When $\Theta$ is complex, the corresponding expression is obtained by deleting the $\Re\{\cdot\}$ operation above.

Consider now the projection (approximation) of $\delta\!\theta$ onto the finite dimensional subspace 
spanned by the orthonormal system $\{\tilde v_i\}_{i=1}^{r}$,
\ie 
\begin{equation}\label{eq:thetaprojdef}
\overline{\delta\!\theta}_r=\sum_{i=1}^{r}\vartheta_i \tilde v_i,
\end{equation}
where $r$ is the dimension of the approximating subspace. 
It follows from (\ref{eq:difforthorel}) and the orthogonality of the noise terms (\ref{eq:covariance1}) and (\ref{eq:covariance2}),
that a sufficient statistics \cite{Kay1993,Scharf1991} for estimating $\overline{\delta\!\theta}_r$ is given by the discrete measurement
model (\ref{eq:suffstat1}), or (\ref{eq:suffstat2}), restricted to the index set $\tilde I_r=\{1,\ldots,r\}$.
The corresponding unbiased estimator is represented as
\begin{equation}\label{eq:unbiasedest}
\widehat{\delta\!\theta}_r=\sum_{i=1}^{r}\widehat\vartheta_i \tilde v_i,
\end{equation}
where $\widehat\vartheta_i$ are unbiased estimates of the principal parameters defined in (\ref {eq:principaldef}).

Assume that $\Theta$ is real. The Cram\'{e}r-Rao lower bound \cite{Kay1993,Scharf1991,VanTrees1968} for estimating the principal parameters 
follows then from the measurement model (\ref{eq:suffstat1}), (\ref{eq:covariance1}) and (\ref{eq:difforthorel}), and is given by
\begin{equation}\label{eq:CRB1}
{\cal E}\{|\widehat{\vartheta_i}-\vartheta_i|^2\}\geq \frac{1}{2\tilde\sigma_i^2}=\frac{1}{\tilde\mu_i},\ i=1,\ldots,r,
\end{equation}
where $\tilde\mu_i=2\tilde\sigma_i^2$ are the eigenvalues of the Fisher information operator defined in (\ref{eq:Fishdef1}).
A similar expression is obtained when $\Theta$ is complex, and $\tilde\mu_i=\tilde\sigma_i^2$.
It follows directly from (\ref{eq:CRB1}) that the mean squared estimation error is lower bounded as
\begin{equation}\label{eq:CRB2}
{\cal E}\{ \|\widehat{\delta\!\theta}_r-\overline{\delta\!\theta}_r\|^2\}=
\sum_{i=1}^{r}{\cal E}\{|\widehat{\vartheta_i}-\vartheta_i|^2\}\geq\sum_{i=1}^r\frac{1}{\tilde\mu_i},
\end{equation}
which is valid for both the real and the complex space $\Theta$. Note that the truncated Fisher information operator  is given by
\begin{equation}\label{eq:Fishertrunc}
{\cal I}_r=\sum_{i=1}^{r}\tilde\mu_i\tilde v_i\langle \tilde v_i,\delta\!\theta \rangle,
\end{equation}
and the corresponding truncated pseudo-inverse
\begin{equation}\label{eq:Fisherptrunc}
{\cal I}_r^+=\sum_{i=1}^{r}\frac{1}{\tilde\mu_i}\tilde v_i\langle \tilde v_i,\delta\!\theta \rangle,
\end{equation}
and the Cram\'{e}r-Rao lower bound \cite{Kay1993,Scharf1991,VanTrees1968} can also be stated as
\begin{equation}
{\cal E}\{\langle \theta_1,\widehat{\delta\!\theta}_r-\overline{\delta\!\theta}_r \rangle\langle \widehat{\delta\!\theta}_r-\overline{\delta\!\theta}_r,\theta_2 \rangle\}
\geq \langle  \theta_1, {\cal I}_r^+\theta_2 \rangle,
\end{equation}
where $\theta_1,\theta_2\in\Theta$. It is immediately observed that the operators ${\cal I}_r$ and ${\cal I}_r^+$ cannot both be trace class in the limit as $r\rightarrow\infty$.
If the infinite-dimensional Fisher information is trace class, then the corresponding pseudo-inverse does not exist, and vice verse.

An interesting interpretation of the Cram\'{e}r-Rao lower bound is obtained as follows.
It is noted that the lower bound defined in (\ref{eq:CRB2}) is a function of the spatial resolution, \ie the subspace dimension $r$.
As the desired spatial resolution $r$ increases, the estimation accuracy decreases
in the sense that the Cram\'{e}r-Rao lower bound increases. The Cram\'{e}r-Rao lower bound
gives the best possible (optimal) performance of any unbiased estimator related to the non-linear inverse
problem at hand. Hence, the lower bound  defined in (\ref{eq:CRB2}) gives
a quantitative interpretation of the optimal trade-off between the accuracy and the resolution
of an inverse problem. 

The truncated pseudo-inverse (\ref{eq:Jpdefrmu}) gives an example of an efficient unbiased estimator of $\overline{\delta\!\theta}_r$ as follows.
Consider a linear version of (\ref{eq:statmodel}) 
\begin{equation}\label{eq:statmodellinearized}
\xi=\psi(\theta_{\rm b})+{\cal J}\delta\!\theta+w,
\end{equation}
where the forward operator $\psi(\theta)$ has been replaced by its first order approximation, the space $\Theta$ is real
and the noise is assumed to have zero mean, \ie ${\cal E}\{w\}=0$.
The corresponding truncated pseudo-inverse is given by
\begin{equation}\label{eq:Jpdefr2}
\widehat{\delta\!\theta}_r=\sum_{i=1}^r\frac{1}{\tilde\sigma_i}\tilde v_i \Re \langle B^{-1}\tilde u_i, \xi-\psi(\theta_{\rm b}) \rangle,
\end{equation}
and the corresponding parameter estimates are hence given by
\begin{equation}\label{eq:parameterest}
\widehat\vartheta_i =\frac{1}{\tilde\sigma_i} \Re \langle B^{-1}\tilde u_i, {\cal J}\delta\!\theta \rangle+\frac{1}{\tilde\sigma_i} \Re \langle B^{-1}\tilde u_i, w \rangle.
\end{equation}
The mean value of (\ref{eq:parameterest}) is given by
\begin{equation}\label{eq:Jpdefr2mean}
\begin{array}{l}
{\cal E}\{\widehat\vartheta_i\}=\displaystyle\frac{1}{\tilde\sigma_i} \Re \langle B^{-1}\tilde u_i, {\cal J}\delta\!\theta \rangle 
 =\frac{1}{\tilde\sigma_i}\langle \tilde u_i, {\cal J}\delta\!\theta \rangle_\mu \vspace{0.2cm}\\
 \hspace{3cm} =\displaystyle\frac{1}{\tilde\sigma_i}\langle {\cal J}_\mu^*\tilde u_i, \delta\!\theta \rangle=\langle \tilde v_i, \delta\!\theta \rangle=\vartheta_i,
\end{array}
\end{equation}
and the variance
\begin{equation}\label{eq:Jpdefr2var}
\textrm{var}\{\widehat\vartheta_i\}=\textrm{var}\{\frac{1}{\tilde\sigma_i} \Re \langle B^{-1}\tilde u_i, w \rangle\}=\frac{1}{2\tilde\sigma_i^2}=\frac{1}{\tilde\mu_i},
\end{equation}
where (\ref{eq:singsystemmu}) and (\ref{eq:covariance1}) have been used. A similar result is obtained when $\Theta$ is complex. Hence, with a linear estimation
model as in (\ref{eq:statmodellinearized}), the truncated pseudo-inverse (\ref{eq:Jpdefrmu}) leads to an unbiased estimator which is efficient in the sense that it
achieves the Cram\'{e}r-Rao lower bound (\ref{eq:CRB2}), see also \cite{Kay1993,Scharf1991,VanTrees1968}. For non-linear estimation problems, 
it should be noted that the Maximum Likelihood (ML) criterion yields an estimation error that approaches the Cram\'{e}r-Rao lower bound 
as the data record is getting large, \ie the ML-estimate is asymptotically efficient, see \cite{Kay1993,Scharf1991,VanTrees1968}.

\subsection{Conditions for the Jacobian}
A sufficient condition is formulated for the Jacobian ${\cal J}$ to be restricted as ${\cal R}\{{\cal J}\}\subset {\cal R}\{B\}$.
This condition also guarantees that ${\cal J}$ is Hilbert-Schmidt with respect to the Cameron-Martin space ${\cal H}_{\mu}$.
Note that the result is valid for both real and complex spaces $\Theta$.

\begin{proposition}\label{prop:proposition1}
Let the Jacobian ${\cal J}:\Theta\rightarrow {\cal H}$ be Hilbert-Schmidt with respect to ${\cal H}$, with the singular system (\ref{eq:singsystem}). 
Let the positive, self-adjoint covariance operator $B:{\cal H}\rightarrow{\cal H}$ be trace class, with the singular system (\ref{eq:Beig}).
Suppose that ${\cal R}\{{\cal J}\}\subset\overline{\textrm{span}\{\phi_j\}_{j\in J}}$.

If
\begin{equation}\label{eq:suffcond2}
\sum_{i\in I}\sigma_i^2\sum_{j\in J}\frac{1}{\lambda_j^2}|\langle\phi_j,u_i \rangle|^2<\infty,
\end{equation}
then ${\cal R}\{{\cal J}\}\subset {\cal R}\{B\}$, and ${\cal J}$ is Hilbert-Schmidt with respect to the Cameron-Martin space ${\cal H}_{\mu}$,
and
\begin{equation}\label{eq:suffcond3}
\textrm{tr}\{{\cal J}_{\mu}^*{\cal J}\}=\sum_{i\in I}\sigma_i^2\sum_{j\in J}\frac{1}{\lambda_j}|\langle\phi_j,u_i \rangle|^2.
\end{equation}
\end{proposition}

{\bf Proof:} For the operator equation $B\xi={\cal J}\theta$ to be solvable, it is necessary and sufficient that
${\cal J}\theta\in {\cal N}\{B\}^\perp$, and that
\begin{equation}\label{eq:ness1}
\sum_{j\in J}\frac{1}{\lambda_j^2}|\langle\phi_j,{\cal J}\theta \rangle|^2<\infty,
\end{equation}
for any $\theta\in\Theta$, see \eg \cite{Kirsch1996}.
Hence, it is assumed that ${\cal R}\{J\}\subset{\cal N}\{B\}^\perp=\overline{\textrm{span}\{\phi_j\}_{j\in J}}$, and it is then sufficient to consider (\ref{eq:ness1}).
By using (\ref{eq:Jdef}), it follows that
\begin{equation}\label{eq:proof1}
\begin{array}{l}
\displaystyle \sum_{j\in J}\frac{1}{\lambda_j^2}|\langle\phi_j,{\cal J}\theta \rangle|^2 \vspace{0.2cm} \\
\hspace{1cm} =\displaystyle\sum_{j\in J}\frac{1}{\lambda_j^2}|\langle\phi_j,\sum_{i\in I}\sigma_iu_i\langle v_i,\theta \rangle \rangle|^2 \vspace{0.2cm} \\
\hspace{1cm} =\displaystyle\sum_{j\in J}\frac{1}{\lambda_j^2}|\sum_{i\in I}\sigma_i\langle\phi_j,u_i\rangle\langle v_i,\theta \rangle |^2 \vspace{0.2cm} \\
\hspace{1cm} \leq\displaystyle\sum_{j\in J}\frac{1}{\lambda_j^2}\left(\sum_{i\in I}\sigma_i^2|\langle\phi_j,u_i\rangle |^2\right)
\left( \sum_{i\in I} |\langle v_i,\theta \rangle|^2 \right) \vspace{0.2cm} \\
\hspace{1cm} =\displaystyle\sum_{j\in J}\frac{1}{\lambda_j^2}\left(\sum_{i\in I}\sigma_i^2|\langle\phi_j,u_i\rangle |^2\right) \|\theta\|^2<\infty,
\end{array}
\end{equation}
for any $\theta\in\Theta$, and where the Cauchy-Schwartz inequality has been used.
Hence, ${\cal R}\{{\cal J}\}\subset {\cal R}\{B\}$.

Now,
\begin{equation}\label{eq:proof2}
\begin{array}{l}
\textrm{tr}\{{\cal J}_{\mu}^*{\cal J}\}=\textrm{tr}\{{\cal J}^*B^{-1}{\cal J}\}\vspace{0.2cm} \\
\hspace{1cm} =\displaystyle\sum_{i\in I}\langle v_i, {\cal J}^*B^{-1}{\cal J}v_i\rangle\vspace{0.2cm} 
=\displaystyle\sum_{i\in I}\langle v_i, {\cal J}^*B^{-1}\sigma_iu_i\rangle\vspace{0.2cm} \\
\hspace{1cm} =\displaystyle\sum_{i\in I}\langle Jv_i, B^{-1}\sigma_iu_i\rangle\vspace{0.2cm} 
=\displaystyle\sum_{i\in I}\langle \sigma_iu_i, B^{-1}\sigma_iu_i\rangle\vspace{0.2cm} \\
\hspace{1cm} =\displaystyle\sum_{i\in I}\sigma_i^2\langle u_i, B^{-1}u_i\rangle,
\end{array}
\end{equation}
and since $u_i\in{\cal R}\{{\cal J}\}\subset{\cal R}\{B\}\subset\overline{\textrm{span}\{\phi_j\}_{j\in J}}$,
\begin{equation}\label{eq:proof3}
\begin{array}{l}
\textrm{tr}\{{\cal J}_{\mu}^*{\cal J}\}=\displaystyle\sum_{i\in I}\sigma_i^2\langle u_i, B^{-1}u_i\rangle\vspace{0.2cm} \\
\hspace{1cm} =\displaystyle\sum_{i\in I}\sigma_i^2\langle u_i, \sum_{j\in J}\frac{1}{\lambda_j}\phi_j\langle \phi_j, u_i \rangle\rangle\vspace{0.2cm} \\
\hspace{1cm} =\displaystyle\sum_{i\in I}\sigma_i^2\sum_{j\in J}\frac{1}{\lambda_j}\langle \phi_j, u_i \rangle\langle u_i, \phi_j\rangle\vspace{0.2cm} \\
\hspace{1cm} =\displaystyle\sum_{i\in I}\sigma_i^2\sum_{j\in J}\frac{1}{\lambda_j}|\langle \phi_j, u_i \rangle|^2<\infty,
\end{array}
\end{equation}
where the last term converges faster than (\ref{eq:suffcond2}) as $\lambda_j\rightarrow 0$. \hfill $\Box$

It is easy to find Hilbert-Schmidt operators ${\cal J}$ and trace class operators $B$ such that the conditions in 
proposition $\ref{prop:proposition1}$ are satisfied.
An example can readily be constructed as follows. Let the Hilbert-Schmidt operator ${\cal J}$ be given, having the singular system (\ref{eq:singsystem}).
It is furthermore assumed that the singular values decay sufficiently fast so that $\sum_{i\in I}\sqrt{\sigma_i}<\infty$.
Let $\phi_i=u_i$ for all $i\in I=J$, and define the operator $B$ by
\begin{equation}
B\xi=\sum_{i\in I}\lambda_i\phi_i\langle\phi_i,\xi \rangle,
\end{equation}
where
\begin{equation}\label{eq:noiseexlambda}
\lambda_i=\left\{
\begin{array}{ll}
\sigma_{\rm w}^2 & i=1,\ldots,q, \vspace{0.2cm} \\
\sqrt{\sigma_i} & i > q,
\end{array}\right.
\end{equation}
and where $\sigma_{\rm w}^2>0$ and $q$ is a positive integer. Note that $B\phi_i=\lambda_i\phi_i$.
Since $\phi_i=u_i$ for all $i\in I=J$ and ${\cal R}\{J\}\subset \overline{\textrm{span}\{u_i\}_{i\in I}}$, 
it follows directly that ${\cal R}\{J\}\subset \overline{\textrm{span}\{\phi_j\}_{j\in J}}$.

The sufficient condition (\ref{eq:suffcond2}) can now be verified as follows:
\begin{equation}\label{eq:versuffcond1}
\begin{array}{l}
\displaystyle\sum_{j\in J}\frac{1}{\lambda_j^2}|\langle\phi_j,u_i \rangle|^2 
=\sum_{j\in J}\frac{1}{\lambda_j^2}|\langle\phi_j,\phi_i \rangle|^2=\frac{1}{\lambda_i^2}<\infty,
\end{array}
\end{equation}
for all $i\in I$, and
\begin{equation}\label{eq:versuffcond2}
\begin{array}{l}
\displaystyle\sum_{i\in I}\sigma_i^2\sum_{j\in J}\frac{1}{\lambda_j^2}|\langle\phi_j,u_i \rangle|^2=\sum_{i\in I}\sigma_i^2\frac{1}{\lambda_i^2} \vspace{0.2cm} \\
\hspace{1cm}=\displaystyle\sum_{i=1}^q\frac{\sigma_i^2}{\sigma_{\rm w}^4}+\sum_{i=q+1}^{\infty}\sigma_i<\infty,
\end{array}
\end{equation}
which shows that ${\cal R}\{{\cal J}\}\subset {\cal R}\{B\}$, and ${\cal J}$ is Hilbert-Schmidt with respect to the Cameron-Martin space ${\cal H}_{\mu}$.

The singular values $\tilde\sigma_i$  defined in (\ref{eq:singsystemmu}) are given by
\begin{equation}
\displaystyle\tilde\sigma_i^2=\frac{\sigma_i^2}{\lambda_i}=\left\{\begin{array}{ll}
\displaystyle\frac{\sigma_i^2}{\sigma_{\rm w}^2} & i=1,\ldots,q, \vspace{0.2cm} \\
\sigma_i^{3/2} & i>q,
\end{array}\right.
\end{equation}
and the singular vectors $\tilde u_i=\sqrt{\lambda_i}u_i$ and $\tilde v_i=v_i$. These relations are
obtained by using ${\cal J}v_i=\sigma_iu_i$, ${\cal J}^*u_i=\sigma_iv_i$, $B^{-1}u_i=\frac{1}{\lambda_i}u_i$, ${\cal J}_{\mu}^*={\cal J}^*B^{-1}$
and $\langle\tilde u_i,\tilde u_j \rangle_{\mu}=\langle \sqrt{\lambda_i}u_i, B^{-1}\sqrt{\lambda_j}u_j\rangle=
\langle \sqrt{\lambda_i}u_i, u_j/\sqrt{\lambda_j}\rangle=\delta_{ij}$.

It is impossible to have a white noise stochastic process
$w$ with a trace class identity covariance operator $B=I$ defined on infinite-dimensional spaces. 
However, as the example above shows, the stochastic process $w$ can be white with identity covariance operator $B=I$ on finite-dimensional subspaces.
In particular, in the example above, let the approximating subspace dimension $r$ be fixed and choose $q>r$ such that 
$\sigma_{r}^2>\sigma_{\rm w}^2\sigma_{q}^{3/2}$, then the first $r$ singular values and singular vectors are given by
\begin{equation}
\left\{\begin{array}{ll}
\tilde\sigma_i^2=\displaystyle\frac{\sigma_i^2}{\sigma_{\rm w}^2} & i=1,\ldots,r, \vspace{0.2cm} \\
\tilde u_i=\sigma_{\rm w}u_i &  i=1,\ldots,r, \vspace{0.2cm} \\
\tilde v_i=v_i &  i=1,\ldots,r,
\end{array}\right.
\end{equation}
and except for a scaling,
the singular systems (\ref{eq:singsystem}) and (\ref{eq:singsystemmu}) are the same for $i=1,\ldots,r$.

\subsection{Unbounded Fisher information and finite Cram\'{e}r-Rao lower bound}\label{sect:unboundedFIM}
It is easy to find Hilbert-Schmidt operators ${\cal J}$ and trace class operators $B$ such that the conditions in 
proposition $\ref{prop:proposition1}$ are not satisfied. In particular, if the condition ${\cal R}\{{\cal J}\}\subset{\cal R}\{{\cal H}_{\mu}\}$ is not satisfied,
then the Cameron-Martin space cannot be used as indicated in (\ref{eq:suffmap}) above.
In this case, it may be difficult (or technically very complicated) to obtain a useful definition of the Fisher information and the Cram\'{e}r-Rao lower bound.
However, a simple case which is also very useful, arises when the singular vectors $u_i$ of the Jacobian operator ${\cal J}$ coincides with the eigenvectors $\phi_j$
of the covariance operator $B$. Hence, the following assumptions are made
\begin{equation}\label{eq:singsystemconcide}
\left\{\begin{array}{l}
{\cal J}v_i=\sigma_i u_i, \vspace{0.2cm}\\
{\cal J}^*u_i=\sigma_i v_i, \vspace{0.2cm}\\
Bu_i=\lambda_iu_i,
\end{array}\right.
\end{equation}
where $i\in I$. For simplicity, it is also assumed here that the space $\Theta$ is complex (the case with real $\Theta$ is similar).

Consider now a sequence of finite-dimensional measurement models based on (\ref{eq:statmodel})
\begin{equation}
P_r\xi=P_r\psi(\theta)+P_rw,
\end{equation}
where $P_r$ denotes the projection operator $P_r:{\cal H}\rightarrow\textrm{span}\{u_1,\ldots,u_r\}$.
The finite-dimensional Jacobian and covariance operators are given by
\begin{equation}
\left\{\begin{array}{l}
P_r{\cal J}\delta\!\theta=\displaystyle\sum_{i=1}^{r}\sigma_iu_i\langle v_i,\delta\!\theta \rangle, \vspace{0.2cm}\\
P_r BP_r \xi=\displaystyle\sum_{i=1}^{r}\lambda_iu_i\langle u_i,\xi \rangle.
\end{array}\right.
\end{equation}
Since the requirement ${\cal R}\{P_r{\cal J}\}\subset{\cal R}\{P_r BP_r\}$ is trivially satisfied, all the results of the previous sections apply in this finite-dimensional case.
In particular, the Fisher information operator (\ref{eq:Fishdef1}) becomes
\begin{equation}\label{eq:Fishdef1finite}
{\cal I}_r\delta\!\theta=\sum_{i=1}^r\frac{\sigma_i^2}{\lambda_i}v_i\langle v_i,\delta\!\theta \rangle,
\end{equation}
and the Cram\'{e}r-Rao lower bound (\ref{eq:CRB2}) is given by
\begin{equation}\label{eq:CRB3}
{\cal E}\{ \|\widehat{\delta\!\theta}_r-\overline{\delta\!\theta}_r\|^2\}\geq\sum_{i=1}^r\frac{\lambda_i}{\sigma_i^2},
\end{equation}
where $\tilde\mu_i=\tilde\sigma_i^2=\sigma_i^2/\lambda_i$, $\tilde u_i=\sqrt{\lambda_i}u_i$ and $\tilde v_i=v_i$,
and where $\overline{\delta\!\theta}_r$ and $\widehat{\delta\!\theta}_r$ are defined as in (\ref{eq:thetaprojdef}) and (\ref{eq:unbiasedest}), respectively.

It is easy to find Hilbert-Schmidt operators ${\cal J}$ and trace class operators $B$ such that the Fisher information operator
(\ref{eq:Fishdef1finite}) does not converge as $r\rightarrow\infty$, but the corresponding pseudo-inverse
\begin{equation}\label{eq:Cramer-Rao}
{\cal I}^+\delta\!\theta=\sum_{i\in I}\frac{\lambda_i}{\sigma_i^2}v_i\langle v_i,\delta\!\theta \rangle,
\end{equation}
is trace class, and the right-hand side of (\ref{eq:CRB3}) converges. It is noted that when $\sigma_i^2/\lambda_i\rightarrow \infty$, it may still be natural
to organize the eigenvalues according to the magnitude of the singular values $\sigma_i$ defined by the Jacobian ${\cal J}$.

The next section will give an example of an inverse problem were the statistical model is given by a physically well-motivated external noise source,
and where the Fisher information operator (\ref{eq:Fishdef1finite}) does not converge, 
the pseudo-inverse (\ref{eq:Cramer-Rao}) is trace class and the Cram\'{e}r-Rao lower bound (\ref{eq:CRB3}) is finite as $r\rightarrow\infty$. 
It should be noted, however,
that in a realistic, real measurement scenario, internal noise in the form of measurement errors are always present. This could mean \eg
the addition of uncorrelated measurement noise. A simple example of this scenario is to consider the addition of the noise eigenvalues
described in (\ref{eq:noiseexlambda}), which will yield a trace class Fisher information (\ref{eq:Fishdef1finite}),
and an infinite Cram\'{e}r-Rao lower bound (\ref{eq:CRB3}).

\section{An electromagnetic inverse source problem with spherically isotropic noise}
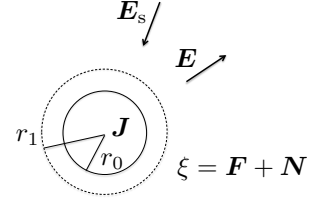
\begin{figure}[htb]
\begin{picture}(50,100)
\put(92,0){\makebox(50,80){\includegraphics[width=3cm]{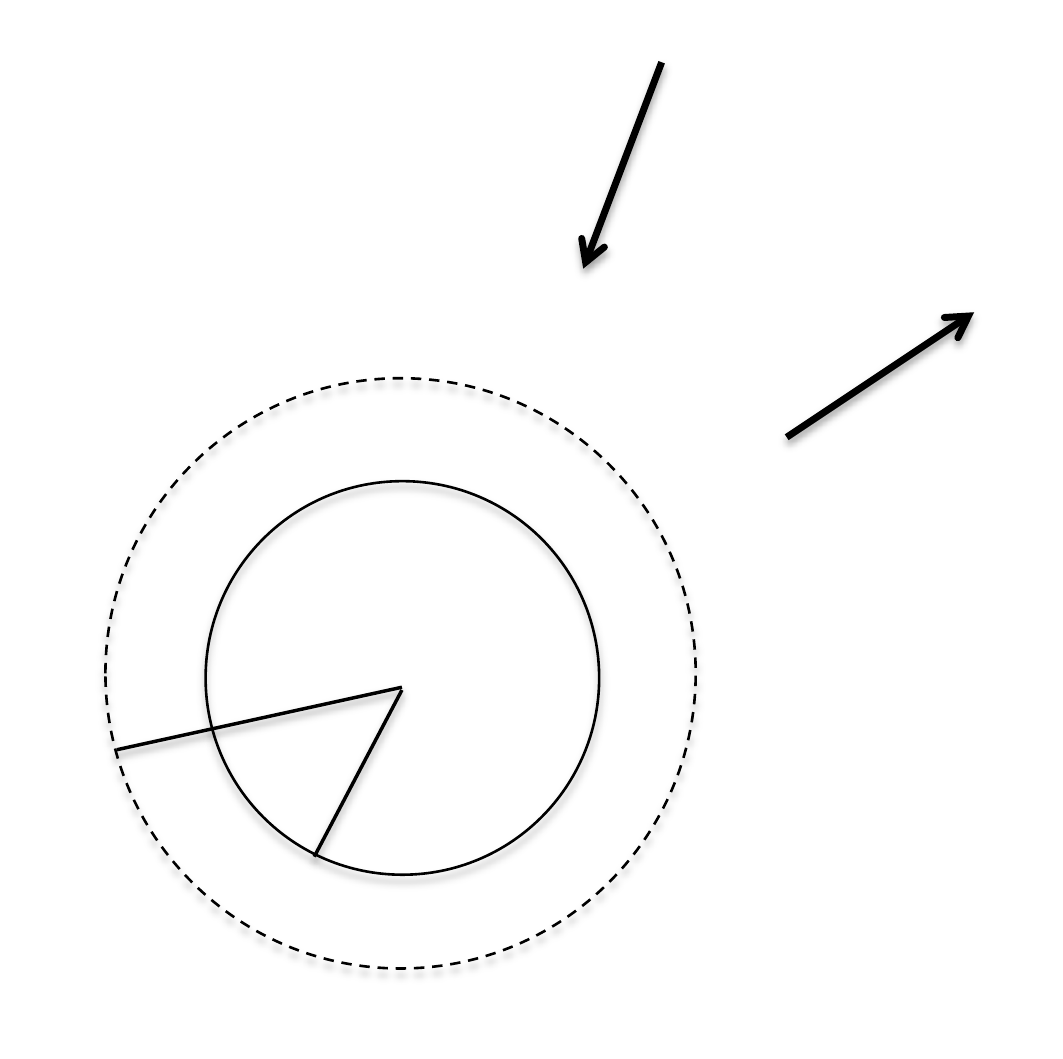}}} 
\put(109,27){$\bm{J}$}
\put(105,16){$r_0$}
\put(73,25){$r_1$}
\put(133,53){$\bm{E}$}
\put(111,71){$\bm{E}_{\rm s}$}
\put(134,12){$\xi=\bm{F}+\bm{N}$}
\end{picture}
\caption{Illustration of the electromagnetic inverse source problem with spherically isotropic noise $\bm{E}_{\rm s}$.}
\label{fig:bild1}
\end{figure}

As an application example of the theory developed in section \ref{sect:CRB}, an electromagnetic inverse source 
problem \cite{Marengo+Devaney1999,Marengo+Ziolkowski2000,Marengo+etal2004,Nordebo+Gustafsson2006}
is considered here, where the observation (or measurement) is corrupted by spherically isotropic noise \cite{Hudson1981,Cron+Sherman1962}, \cf Fig.\ \ref{fig:bild1}.
Here, $\bm{J}$ is the current source which is contained within a sphere of radius $r_0$, $\bm{E}$ is the transmitted electric field and $\bm{E}_{\rm s}$ the spherically isotropic noise.
The tangential components of the fields are observed at a sphere of radius $r_1$. 
The objective here is to quantify the optimal estimation performance of the truncated pseudo-inverse (\ref{eq:Jpdefrmu})
in terms of the Cram\'{e}r-Rao lower bound (\ref{eq:CRB2}), where $\delta\!\theta=\bm{J}$.

Below, $r$, $\theta$ and $\phi$ will denote the spherical coordinates, and $\bm{r}=r\hat{\bm{r}}$ the radius vector where $\hat{\bm{r}}$ is the corresponding unit vector.
The time convention is defined by the factor $\eu^{-\iu \omega t}$ where $\omega$ is the angular frequency and $t$ the time.
Furthermore, let $k$, ${\rm c_0}$ and $\eta_0$ denote the wave number, the speed of light and the wave impedance of free space, respectively, where $k=\omega/{\rm c}_0$.

The transmitted electric field $\bm{E}(\bm{r})$ satisfies the Maxwell's equations \cite{Jackson1999} and the following vector wave equation
\begin{equation}\label{eq:vecwaveeq}
\nabla\times\nabla\times\bm{E}(\bm{r})-k^2\bm{E}(\bm{r})=\iu k\eta_0\bm{J}(\bm{r}),
\end{equation}
and the spherically isotropic noise ${\bm E}_{\rm s}(\bm{r})$ satisfies similarly the corresponding source-free Maxwell's equations.
The observation model (\ref{eq:statmodel}) is given by
\begin{equation}
\xi=\bm{F}(\hat{\bm{r}})+\bm{N}(\hat{\bm{r}}),
\end{equation}
where $\bm{F}(\hat{\bm{r}})=-\hat{\bm r}\times\hat{\bm r}\times\bm{E}(r_1\hat{\bm{r}})$ and $\bm{N}(\hat{\bm{r}})=-\hat{\bm r}\times\hat{\bm r}\times\bm{E}_{\rm s}(r_1\hat{\bm{r}})$ 
are the tangential components of the transmitted electric field $\bm{E}(\bm{r})$ and the noise field $\bm{E}_{\rm s}(\bm{r})$, respectively,
as they are observed at a sphere of radius $r_1$. 

The space $\Theta$ consists of all complex vector fields $\bm{J}(\bm{r})$ which are square integrable over the spherical volume $V_{r_0}$ with radius $r_0$, 
and is equipped with the scalar product
\begin{equation}\label{eq:scaldefJ}
\langle \bm{J}_1,\bm{J}_2 \rangle=\int_{V_{r_0}}\bm{J}_1^*(\bm{r})\cdot\bm{J}_2(\bm{r}){\rm d}v,
\end{equation}
where $\{\cdot\}^*$ denotes the complex conjugate. The space ${\cal H}$ consists of all tangential vector fields $\bm{F}$ (with $\hat{\bm{r}}\cdot\bm{F}=0$)
which are square integrable over the spherical surface $S_{r_1}$ with radius $r_1$, and is equipped with the scalar product
\begin{equation}\label{eq:scaldefF}
\langle \bm{F}_1,\bm{F}_2 \rangle=\int_{S_{r_1}}\bm{F}_1^*(\hat{\bm{r}})\cdot\bm{F}_2(\hat{\bm{r}}){\rm d}S.
\end{equation}

\subsection{Forward model based on vector spherical waves}
The electromagnetic fields in a source-free region can be expressed in terms of the vector spherical waves defined in appendix \ref{sect:spherical},
see also \cite{Bostrom+Kristensson+Strom1991,Arfken+Weber2001,Jackson1999,Newton2002}.
Hence, by considering the free-space Green's dyadic defined in (\ref{eq:Greensdyadic2}), the forward operator
${\cal J}:\Theta\rightarrow{\cal H}$ is given by
\begin{equation}\label{eq:Jdefsph}
\begin{array}{l}
{\cal J}\bm{J}(\bm{r})=-k^2\eta_0\displaystyle\sum_{\tau=1}^2\sum_{l=1}^{\infty}\sum_{m=-l}^{l}f_{\tau lm}\bm{A}_{\tau lm}(\hat{\bm{r}})  \vspace{0.2cm}\\
\hspace{2cm} \displaystyle \int_{V_{r_0}}\bm{v}_{\tau lm}^\dagger(k\bm{r}^\prime)\cdot \bm{J}(\bm{r}^\prime){\rm d}v^\prime,
\end{array}
\end{equation}
where $\bm{A}_{\tau lm}(\hat{\bm{r}})$ are the vector spherical harmonics and $\bm{v}_{\tau lm}(k\bm{r})$ the regular vector spherical waves 
defined in (\ref{eq:udef}) and (\ref{eq:Adef}).  Note that the dagger notation $(\cdot)^\dagger$ is also defined in appendix \ref{sect:spherical}.
The coefficients $f_{\tau lm}$ are given by
\begin{equation}\label{eq:ftaulmdef}
\begin{array}{l}
f_{1lm}={\rm h}_l^{(1)}(kr_1), \vspace{0.2cm}\\
f_{2lm}= \displaystyle\frac{(kr_1{\rm h}_l^{(1)}(kr_1))^\prime}{kr_1},
\end{array}
\end{equation}
where ${\rm h}_l^{(1)}(x)$ are the spherical Hankel functions of the first kind.
It is readily seen that the operator ${\cal J}$ has a non-empty nullspace (non-radiating sources).
In particular, by writing $\bm{v}_{2 ml}(k\bm{r})=a(r)\bm{A}_{2ml}(\hat{\bm{r}})+b(r)\bm{A}_{3ml}(\hat{\bm{r}})$, it follows that
${\cal J}\left(-b(r)\bm{A}_{2ml}(\hat{\bm{r}})+a(r)\bm{A}_{3ml}(\hat{\bm{r}})\right)=0$, where the orthonormality of the vector spherical harmonics (\ref{eq:Aorthonormal})
has been used.

The adjoint operator ${\cal J}^*:{\cal H}\rightarrow\Theta$ is readily obtained as
\begin{equation}\label{eq:Jadjdefsph}
\begin{array}{l}
{\cal J}^*\bm{F}(\hat{\bm{r}})=-k^2\eta_0\displaystyle\sum_{\tau=1}^2\sum_{l=1}^{\infty}\sum_{m=-l}^{l}\bm{v}_{\tau lm}(k\bm{r})f_{\tau lm}^* \vspace{0.2cm}\\
\hspace{2cm} \displaystyle \int_{S_{r_1}}\bm{A}_{\tau lm}^*(\hat{\bm{r}}^\prime)\cdot \bm{F}(\hat{\bm{r}}^\prime){\rm d}S^\prime,
\end{array}
\end{equation}
where $r\leq r_0$, so that $\langle {\cal J}\bm{J}(\bm{r}), \bm{F}(\hat{\bm{r}})\rangle=\langle \bm{J}(\bm{r}), {\cal J}^*\bm{F}(\hat{\bm{r}})\rangle$.
The combined, self-adjoint operator ${\cal J}{\cal J}^*:{\cal H}\rightarrow{\cal H}$ is then given by
\begin{equation}\label{eq:JJadjdefsph}
\begin{array}{l}
{\cal J}{\cal J}^*\bm{F}(\hat{\bm{r}})=k^4\eta_0^2\displaystyle\sum_{\tau=1}^2\sum_{l=1}^{\infty}\sum_{m=-l}^{l}|f_{\tau lm}|^2\int_{V_{r_0}}|\bm{v}_{\tau lm}(k\bm{r})|^2{\rm d}v\vspace{0.2cm}\\
\hspace{2cm} \displaystyle\bm{A}_{\tau lm}(\hat{\bm{r}})\int_{S_{r_1}}\bm{A}_{\tau lm}^*(\hat{\bm{r}}^\prime)\cdot\bm{F}(\hat{\bm{r}}^\prime){\rm d}S^\prime,
\end{array}
\end{equation}
where the orthogonality of the regular vector spherical waves (\ref{eq:vorthogonal}) has been used. It follows immediately from 
(\ref{eq:JJadjdefsph}) that the vector spherical harmonics $\bm{A}_{\tau lm}(\hat{\bm{r}})$ 
are eigenvectors, and that the eigenvalues (the squared singular values) are given by
\begin{equation}\label{eq:eigJJadj1}
\sigma_{\tau lm}^2=k^4\eta_0^2r_1^2 |f_{\tau lm}|^2\bar\sigma_{\tau lm}^2,
\end{equation}
where
\begin{equation}\label{eq:eigJJadj2}
\bar\sigma_{\tau lm}^2=\int_{V_{r_0}}|\bm{v}_{\tau lm}(k\bm{r})|^2{\rm d}v.
\end{equation}
The eigenvectors of the operator ${\cal J}^*{\cal J}$ are similarly given by the regular vector spherical waves $\bm{v}_{\tau lm}(k\bm{r})$ for $r\leq r_0$,
and the corresponding eigenvalues are again given by (\ref{eq:eigJJadj1}). 

The factor (\ref{eq:eigJJadj2}) can be evaluated explicitly as
\begin{equation}
\begin{array}{l}
\bar\sigma_{1 lm}^2=\displaystyle\frac{r_0^3}{2}({\rm j}_l^2(kr_0)-{\rm j}_{l-1}(kr_0){\rm j}_{l+1}(kr_0)), \vspace{0.2cm} \\
\bar\sigma_{2 lm}^2=\displaystyle\frac{1}{2l+1}\left((l+1)\bar\sigma_{1 (l-1)m}^2+l\bar\sigma_{1 (l+1)m}^2 \right),
\end{array}
\end{equation}
where the relation ${\rm j}_l(kr)=\sqrt{\pi/(2kr)}{\rm J}_{l+1/2}(kr)$ (where $J_{\nu}(\cdot)$ is the Bessel function of order $\nu$) has been used together
with the second Lommel integral \cite{Marengo+Devaney1999,Arfken+Weber2001}, as well as the
recurrence relations for the spherical Bessel functions \cite{Olver+etal2010}. Note that the eigenvalues in (\ref{eq:eigJJadj1}) are independent of the $m$-index.

The asymptotic behavior of the eigenvalues $\sigma_{\tau lm}^2$ in (\ref{eq:eigJJadj1}) for large values of $l$, can be analyzed by using
\begin{equation}\label{eq:asssphbessel}
\begin{array}{l}
{\rm j}_l(kr)\sim\displaystyle \frac{1}{\sqrt{4kr}}\frac{1}{\sqrt{l}}\left(\frac{\eu kr /2}{l}\right)^l, \vspace{0.2cm} \\
{\rm h}_l^{(1)}(kr)\sim\displaystyle -\iu\frac{1}{\sqrt{kr}}\frac{1}{\sqrt{l}}\left(\frac{\eu kr /2}{l}\right)^{-l},
\end{array}
\end{equation}
see \cite{Olver+etal2010}. After some algebra, it is concluded that
\begin{equation}
\bar\sigma_{\tau lm}^2=\left(\frac{\eu kr_0/2}{l} \right)^{2l}o(\frac{1}{l}),
\end{equation}
where $o(\cdot)$ denotes the little ordo \cite{Olver1997}, and hence
\begin{equation}\label{eq:asssigmataulm}
\begin{array}{l}
\sigma_{\tau lm}^2=\displaystyle k^4\eta_0^2r_1^2 |f_{\tau lm}|^2\bar\sigma_{\tau lm}^2 \vspace{0.2cm} \\
\hspace{1cm}=\displaystyle\frac{1}{l}\left(\frac{\eu kr_1 /2}{l}\right)^{-2l}\left(\frac{\eu kr_0/2}{l} \right)^{2l}o(\frac{1}{l})\vspace{0.2cm} \\
\hspace{2cm}=\displaystyle\left(\frac{r_0}{r_1} \right)^{2l}o(\frac{1}{l^2}).
\end{array}
\end{equation}
It is concluded that $\sigma_{\tau lm}^2\rightarrow 0$ as $l\rightarrow\infty$, and the convergence is exponential.
Hence, the operator ${\cal J}$ is Hilbert-Schmidt.

\subsection{Spherically isotropic noise}
Spherically isotropic noise \cite{Hudson1981,Cron+Sherman1962} models a situation where the receiving sensors (antennas, microphones, etc.) 
are subjected to external noise consisting of plane waves impinging from arbitrary directions and with uncorrelated amplitudes.
The following electromagnetic model will be used here to model the spherically isotropic noise
\begin{equation}
\bm{E}_{\rm s}(\bm{r})=\frac{1}{4\pi}\int_{\Omega}\bm{E}_0(\hat{\bm{k}})\eu^{\iu k\hat{\bm{k}}\cdot\bm{r}}{\rm d}\Omega(\hat{\bm{k}}),
\end{equation}
where $\Omega$ is the unit sphere, $\hat{\bm{k}}$ the unit wave vector of the plane partial waves 
$\bm{E}_{\rm s}(\bm{r},\hat{\bm{k}})=\bm{E}_0(\hat{\bm{k}})\eu^{\iu k\hat{\bm{k}}\cdot\bm{r}}$, and ${\rm d}\Omega(\hat{\bm{k}})$ the differential solid angle.
Here, $\bm{E}_0(\hat{\bm{k}})=E_0(\alpha_1(\hat{\bm{k}})\hat{\bm{e}}_1(\hat{\bm{k}})+\alpha_2(\hat{\bm{k}})\hat{\bm{e}}_2(\hat{\bm{k}}))$ 
is modeled as a white (uncorrelated) zero mean complex Gaussian stochastic process in the variable $\hat{\bm{k}}$,
where $E_0$ is a constant and $\hat{\bm{k}}\cdot\bm{E}_0(\hat{\bm{k}})=0$. Here,
$\hat{\bm{k}}$, $\hat{\bm{e}}_1(\hat{\bm{k}})$ and $\hat{\bm{e}}_2(\hat{\bm{k}})$ are the unit vectors in the spherical coordinate system and $\alpha_1(\hat{\bm{k}})$
and $\alpha_2(\hat{\bm{k}})$ the corresponding components of $\bm{E}_0(\hat{\bm{k}})$. The covariance dyadic of $\bm{E}_0(\hat{\bm{k}})$ is given by
\begin{equation}\label{eq:covariancedyadic}
{\cal E}\{\bm{E}_0(\hat{\bm{k}})\bm{E}_0^*(\hat{\bm{k}}^\prime)\}=E_0^2\bm{I}_{2\times 2}(\hat{\bm{k}})\delta(\hat{\bm{k}}-\hat{\bm{k}}^\prime),
\end{equation}
where $\bm{I}_{2\times 2}(\hat{\bm{k}})=-\hat{\bm{k}}\times\hat{\bm{k}}\times$ is the projection dyadic perpendicular to $\hat{\bm{k}}$.

The plane partial waves $\bm{E}_{\rm s}(\bm{r},\hat{\bm{k}})$ can be expanded in regular vector spherical waves as
\begin{equation}\label{eq:planeexp}
\bm{E}_0(\hat{\bm{k}})\eu^{\iu k\hat{\bm{k}}\cdot\bm{r}}
=\displaystyle\sum_{\tau=1}^2\sum_{l=1}^{\infty}\sum_{m=-l}^{l}a_{\tau lm}(\hat{\bm{k}})\bm{v}_{\tau lm}(k\bm{r}),
\end{equation}
where the stochastic expansion coefficients are given by
\begin{equation}
a_{\tau lm}(\hat{\bm{k}})=4\pi\iu^{l-\tau-1}\bm{A}_{\tau lm}^*(\hat{\bm{k}})\cdot\bm{E}_0(\hat{\bm{k}}),
\end{equation}
see \eg \cite{Bostrom+Kristensson+Strom1991}.
By exploiting the plane wave expansion (\ref{eq:planeexp}), the covariance dyadic (\ref{eq:covariancedyadic}) as well as 
the orthonormality of the vector spherical harmonics (\ref{eq:Aorthonormal}), it is readily seen that the covariance dyadic of
the spherically isotropic noise $\bm{E}_{\rm s}(\bm{r})$ is given by
\begin{equation}\label{eq:covariancedyadicEs1}
{\cal E}\{\bm{E}_{\rm s}(\bm{r})\bm{E}_{\rm s}^*(\bm{r}^\prime)\}=E_0^2\sum_{\tau=1}^2\sum_{l=1}^{\infty}\sum_{m=-l}^{l}\bm{v}_{\tau lm}(k\bm{r})\bm{v}_{\tau lm}^*(k\bm{r}^\prime).
\end{equation}

By using $\bm{v}_{\tau lm}=(\bm{u}_{\tau lm}+\bm{w}_{\tau lm})/2$, the dagger relations (\ref{eq:daggerrel}) as well as
the expressions for the free-space Green's dyadic  (\ref{eq:Greensdyadic1}) and (\ref{eq:Greensdyadic2}), it can be shown that
the covariance dyadic (\ref{eq:covariancedyadicEs1}) of the spherically isotropic noise is also given by
\begin{equation}\label{eq:covariancedyadicEs2}
\begin{array}{l}
{\cal E}\{\bm{E}_{\rm s}(\bm{r})\bm{E}_{\rm s}^*(\bm{r}^\prime)\}=\displaystyle E_0^2\frac{1}{k}\Im\{ \bm{G}_{\rm e}(k,\bm{r},\bm{r}^\prime)\} \vspace{0.2cm}\\
=\displaystyle E_0^2\frac{1}{4\pi}\left({\bm I}+\frac{1}{k^2}\nabla\nabla \right)\frac{\sin k|\bm{r}-\bm{r}^\prime|}{k|\bm{r}-\bm{r}^\prime|}.
\end{array}
\end{equation}

The observed noise field is defined here by
\begin{equation}
\bm{N}(\hat{\bm{r}})=-\hat{\bm{r}}\times\hat{\bm{r}}\times\bm{E}_{\rm s}(r_1\hat{\bm{r}})=\bm{I}_{2\times 2}(\hat{\bm{r}})\cdot\bm{E}_{\rm s}(r_1\hat{\bm{r}}),
\end{equation}
and the corresponding covariance dyadic is hence given by
\begin{equation}\label{eq:covariancedyadicN1}
\begin{array}{l}
{\cal E}\{\bm{N}(\hat{\bm{r}})\bm{N}^*(\hat{\bm{r}}^\prime)\}\vspace{0.2cm}\\
\hspace{0.5cm}=\bm{I}_{2\times 2}(\hat{\bm{r}}){\cal E}\{\bm{E}_{\rm s}(r_1\hat{\bm{r}})\bm{E}_{\rm s}^*(r_1\hat{\bm{r}}^\prime)\}\bm{I}_{2\times 2}(\hat{\bm{r}}^\prime)\vspace{0.2cm}\\
\hspace{1cm}=\displaystyle E_0^2\sum_{\tau=1}^2\sum_{l=1}^{\infty}\sum_{m=-l}^{l}g_{\tau lm}^2\bm{A}_{\tau lm}(\hat{\bm{r}})\bm{A}_{\tau lm}^*(\hat{\bm{r}}^\prime)
\end{array}
\end{equation}
where
\begin{equation}\label{eq:gtaulmdef}
\begin{array}{l}
g_{1lm}={\rm j}_l(kr_1), \vspace{0.2cm}\\
g_{2lm}= \displaystyle \displaystyle\frac{(kr_1{\rm j}_l(kr_1))^\prime}{kr_1},
\end{array}
\end{equation}
where ${\rm j}_l(x)$ are the spherical Bessel functions, and where (\ref{eq:covariancedyadicEs1}) and (\ref{eq:udef}) have been used. 
The covariance operator $B:{\cal H}\rightarrow{\cal H}$ for the Gaussian vector $\bm{N}(\hat{\bm{r}})$ is 
defined by the property ${\cal E}\{\langle \bm{F}_1(\hat{\bm{r}}),\bm{N}(\hat{\bm{r}})\rangle\langle \bm{F}_2(\hat{\bm{r}}),\bm{N}(\hat{\bm{r}})\rangle^*\}=
\langle \bm{F}_1(\hat{\bm{r}}),B\bm{F}_2(\hat{\bm{r}})\rangle$, and is obtained as
\begin{equation}\label{eq:BFdef}
\begin{array}{l}
B\bm{F}(\hat{\bm{r}})=\displaystyle E_0^2\sum_{\tau=1}^2\sum_{l=1}^{\infty}\sum_{m=-l}^{l}g_{\tau lm}^2\bm{A}_{\tau lm}(\hat{\bm{r}})\vspace{0.2cm}\\
\hspace{3cm}\displaystyle\int_{S_{r_1}}\bm{A}_{\tau lm}^*(\hat{\bm{r}}^\prime)\cdot \bm{F}(\hat{\bm{r}}^\prime){\rm d}S^\prime.
\end{array}
\end{equation}
It follows immediately from (\ref{eq:BFdef}) that the vector spherical harmonics $\bm{A}_{\tau lm}(\hat{\bm{r}})$ 
are eigenvectors, and that the eigenvalues are given by
\begin{equation}\label{eq:eigBdef}
\lambda_{\tau lm}=E_0^2r_1^2 g_{\tau lm}^2.
\end{equation}
Note that the eigenvalues in (\ref{eq:eigBdef}) are idependent of the $m$-index.
The asymptotic behavior of the eigenvalues $\lambda_{\tau lm}$ for large values of $l$ can be analyzed by using
(\ref{eq:asssphbessel}), which yields
\begin{equation}\label{eq:asslambdataulm}
\lambda_{\tau lm}\sim b_\tau\frac{1}{l}\left(\frac{\eu kr_1/2}{l} \right)^{2l},
\end{equation}
where $b_\tau$ is a constant. It is concluded that $\lambda_{\tau lm}\rightarrow 0$ as $l\rightarrow\infty$, and the convergence is faster than
exponential. Hence, the covariance operator $B$ is trace class.

\subsection{Fisher information and the Cram\'{e}r-Rao lower bound}
It is observed that the singular vectors $u_i$ of the forward operator ${\cal J}$, and the eigenvectors $\phi_j$ of the covariance operator $B$ 
coincide here with the vector spherical harmonics $\bm{A}_{\tau lm}(\hat{\bm{r}})$. Hence, the situation is as described in
section \ref{sect:unboundedFIM} above.
The Cram\'{e}r-Rao lower bound (\ref{eq:CRB3}) is given by
\begin{equation}\label{eq:CRBL}
\textrm{CRB}(L)=\sum_{\tau=1}^2\sum_{l=1}^{L}\sum_{m=-l}^{l}\frac{\lambda_{\tau lm}}{\sigma_{\tau lm}^2},
\end{equation}
where the eigenvalues are organized according to increasing $l$-index (multipole order).

The asymptotics of the singular values $\sigma_{\tau lm}^2$ of the forward operator ${\cal J}$, and the eigenvalues $\lambda_{\tau lm}$ of the
covariance operator $B$ have been given in (\ref{eq:asssigmataulm}) and (\ref{eq:asslambdataulm}) above, respectively.
The asymptotics of the eigenvalues of the Fisher information (\ref{eq:Fishdef1finite}) is hence given by 
\begin{equation}
\frac{\sigma_{\tau lm}^2}{\lambda_{\tau lm}}=\left(\frac{r_0}{r_1} \right)^{2l}\left(\frac{l}{\eu kr_1/2} \right)^{2l}o(\frac{1}{l}),
\end{equation}
which implies that $\sigma_{\tau lm}^2/\lambda_{\tau lm}\rightarrow\infty$, and the Fisher information does not converge.
However, the Cram\'{e}r-Rao lower bound (\ref{eq:CRBL}) converges to a finite value as $L\rightarrow\infty$.

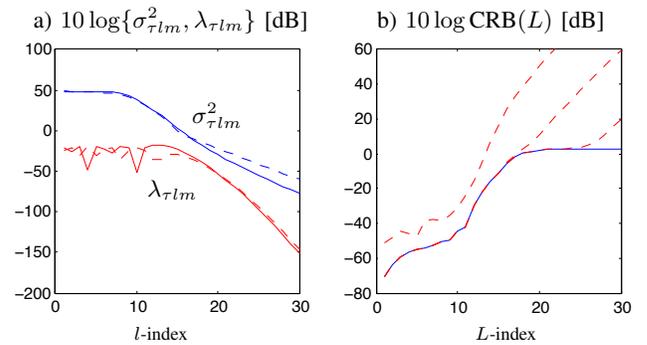
\begin{figure}[htb]
\begin{picture}(50,150)
\put(102,0){\makebox(50,120){\includegraphics[width=8.8cm]{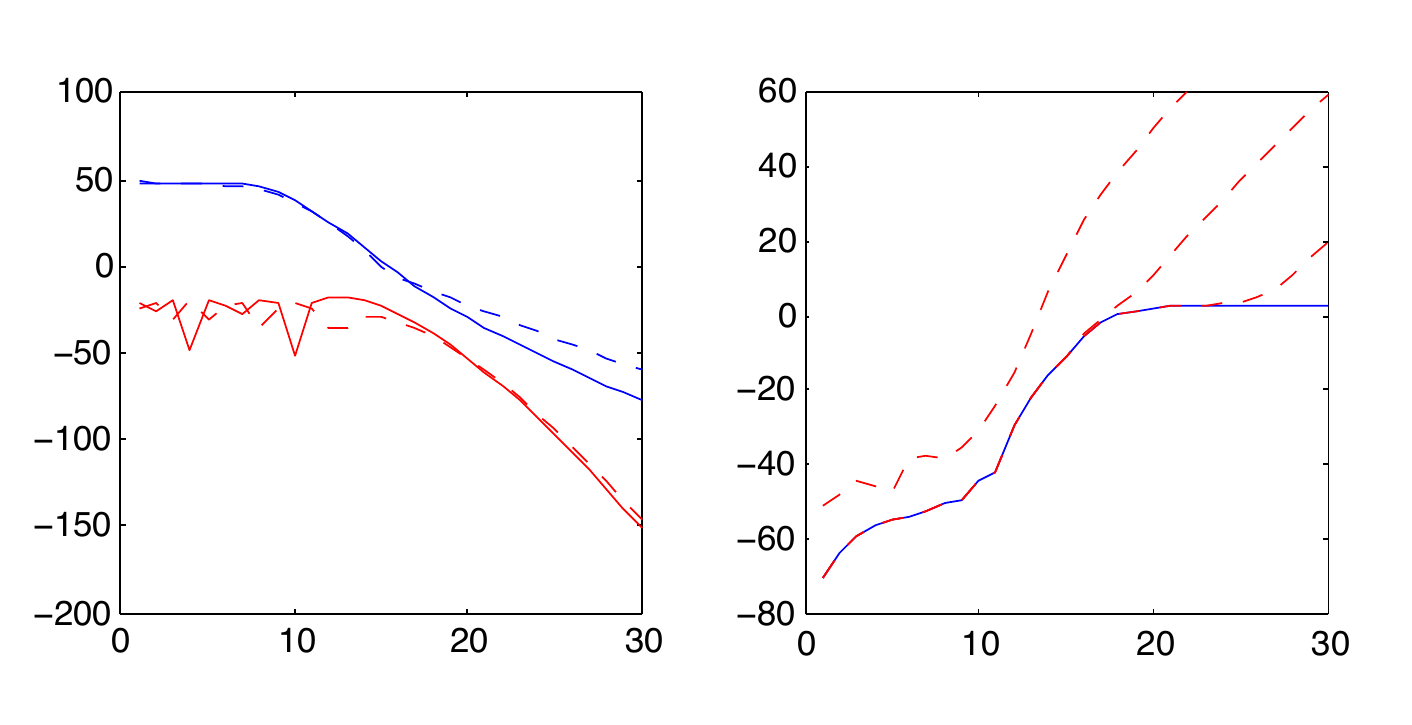}}} 
\put(15,116){\small a) $10\log\{\sigma^2_{\tau lm},\lambda_{\tau lm}\}$ [dB]}
\put(145,116){\small b) $10\log \textrm{CRB}(L)$ [dB]}
\put(75,80){\small $\sigma^2_{\tau lm}$}
\put(58,52){\small $\lambda_{\tau lm}$}
\put(53,-2){\scriptsize $l$-index}
\put(182,-2){\scriptsize $L$-index}
\end{picture}
\caption{a) The singular values $\sigma^2_{\tau lm}$ of the forward operator (upper blue lines) 
and the eigenvalues $\lambda_{\tau lm}$ of the noise covariance (lower red lines). The solid and dashed lines correspond to $\tau=1$ and $\tau=2$, respectively.
b) The Cram\'{e}r-Rao lower bound $\textrm{CRB}(L)$ for the electromagnetic inverse source problem with spherically isotropic noise.
The solid (blue) line corresponds to spherically isotropic noise only, and the dashed (red) lines correspond to white noise added with 
(from lower to upper dashed line) $10\log\textrm{WNR}=-60,-20,20$ \unit{dB}.}
\label{fig:matfig1}
\end{figure}

In Fig.\ \ref{fig:matfig1} a) is illustrated the convergence of the eigenvalues $\sigma^2_{\tau lm}$ 
and $\lambda_{\tau lm}$ given by (\ref{eq:eigJJadj1}) and (\ref{eq:eigBdef}), respectively,
and in Fig.\ \ref{fig:matfig1} b) (the solid line) the convergence of the Cram\'{e}r-Rao lower bound (\ref{eq:CRBL}).
Here, $kr_0=10$, $r_0=1$, $r_1=1.5$ and $E_0=1$.

It is emphasized that the spherically isotropic noise corresponds to an ``external'' noise source.
Suppose now that ``internal'' measurement errors are also present in the form of 
uncorrelated measurement noise with variance $\sigma_{\rm w}^2$, which is added to the eigenvalues $\lambda_{\tau lm}$,
see also the noise model described in (\ref{eq:noiseexlambda}).
The white-noise-ratio (WNR) is defined here by
 \begin{equation}
 \textrm{WNR}=\displaystyle\frac{\sigma_{\rm w}^2}{\displaystyle\max_{\tau lm}\lambda_{\tau lm}},
 \end{equation}
 and the noise eigenvalues in (\ref{eq:CRBL}) are then replaced as $\lambda_{\tau lm}\rightarrow \lambda_{\tau lm}+\sigma_{\rm w}^2$.
 In this case, the Fisher information (\ref{eq:Fishdef1finite}) converges, and the Cram\'{e}r-Rao lower bound (\ref{eq:CRBL}) becomes infinite as $L\rightarrow\infty$.
 The situation is illustrated in Fig.\ \ref{fig:matfig1} b) (the dashed lines) where $10\log\textrm{WNR}=-60,-20,20$ \unit{dB}.
It is clear that if the white noise level is significantly lower than the maximum eigenvalue of the spherically isotropic noise, \ie if
$\sigma_{\rm w}^2<<\max_{\tau lm}\lambda_{\tau lm}$, then it is essential to incorporate the eigenvalues related to the spherically isotropic noise into the
Fisher information analysis, and not only the singular values $\sigma_{\tau lm}$ of the Jacobian ${\cal J}$.

\section{Summary and conclusions}
It is natural to consider an inverse imaging problem as an infinite-dimensional estimation problem based on a statistical observation model.
With Gaussian noise on infinite-dimensional Hilbert space, a trace class covariance operator and a Hilbert-Schmidt Jacobian,
the appropriate space for defining the Fisher information operator is given by the Cameron-Martin space.
A sufficient condition is given for the existence of a trace class Fisher information, which is based solely on the
spectral properties of the covariance operator and of the Jacobian, respectively.
Two important special cases arises: 
1) The infinite-dimensional Fisher information operator exists and is trace class, and
the corresponding pseudo-inverse (and the Cram\'{e}r-Rao lower bound) exists only for finite-dimensional subspaces.
2) The infinite-dimensional pseudo-inverse (and the Cram\'{e}r-Rao lower bound) exists, and the corresponding
Fisher information operator exists only for finite-dimensional subspaces.
An explicit example is given regarding an electromagnetic inverse source problem with ``external'' spherically isotropic noise,
as well as ``internal'' additive uncorrelated noise.

\appendix
\section{Vector spherical waves}\label{sect:spherical}
The regular vector spherical waves are defined here by
\begin{equation}\label{eq:udef}
\begin{array}{l}
\displaystyle\bm{v}_{1 lm}(k{\bm{r}})  =   \frac{1}{\sqrt{l(l+1)}}\nabla\times({\bm{r}}{\rm j}_l(kr)Y_{lm}(\hat{\bm{r}})) \vspace{0.2cm}\\
=   {\rm j}_l(kr)\bm{A}_{1 lm}(\hat{\bm{r}}), \vspace{0.2cm}\\
\bm{v}_{2 lm}(k\bm{r})   =   \displaystyle \frac{1}{k}\nabla\times\bm{v}_{1 lm}(k\bm{r})\vspace{0.2cm}  \\
 =\displaystyle\frac{(kr{\rm j}_l(kr))^{\prime}}{kr}\bm{A}_{2 lm}(\hat{\bm{r}})+\sqrt{l(l+1)}\frac{{\rm j}_l(kr)}{kr}\bm{A}_{3 lm}(\hat{\bm{r}}), 
\end{array}
\end{equation}
where $\bm{A}_{\tau lm}(\hat{\bm{r}})$ are the vector spherical harmonics and ${\rm j}_l(x)$ the spherical Bessel functions,
\cf \cite{Bostrom+Kristensson+Strom1991,Arfken+Weber2001,Jackson1999,Newton2002,Olver+etal2010}. The indices are given by $l=1,\ldots,\infty$ and $m=-l,\ldots,l$.
The vector spherical harmonics $\bm{A}_{\tau lm}(\hat{\bm{r}})$ are given by
\begin{equation}\label{eq:Adef}
\begin{array}{l}
\bm{A}_{1lm}(\hat{\bm{r}})  =   \displaystyle\frac{1}{\sqrt{l(l+1)}}\nabla\times\left( \bm{r}{\rm Y}_{lm}(\hat{\bm{r}}) \right), \vspace{0.2cm}\\
\bm{A}_{2lm}(\hat{\bm{r}})  =  \hat{\bm{r}}\times\bm{A}_{1lm}(\hat{\bm{r}}), \vspace{0.2cm}\\
\bm{A}_{3lm}(\hat{\bm{r}}) = \hat{\bm{r}}{\rm Y}_{lm}(\hat{\bm{r}}),
\end{array}
\end{equation}
where ${\rm Y}_{lm}(\hat{\bm{r}})$ are the scalar spherical harmonics given by
\begin{equation}
{\rm Y}_{lm}(\theta,\phi)=(-1)^m\sqrt{\frac{2l+1}{4\pi}}\sqrt{\frac{(l-m)!}{(l+m)!}}{\rm P}_{l}^m(\cos\theta)\eu^{{\rm i}m\phi},
\end{equation}
and where ${\rm P}_{l}^m(x)$ are the associated Legendre functions \cite{Arfken+Weber2001}.

The vector spherical harmonics are orthonormal on the unit sphere, and hence
\begin{equation}\label{eq:Aorthonormal}
\int_{S_1}\bm{A}_{\tau lm}^*(\hat{\bm{r}})\cdot\bm{A}_{\tau^\prime l^\prime m^\prime}(\hat{\bm{r}}){\rm d}\Omega=\delta_{\tau\tau^\prime}\delta_{ll^\prime}\delta_{mm^\prime},
\end{equation}
where $S_1$ denotes the unit sphere, ${\rm d}\Omega=\sin\!\theta{\rm d}\theta{\rm d}\phi$ and $\tau=1,2,3$. 
As a consequence, the regular vector spherical waves are orthogonal over a spherical volume $V_{r_0}$ with
\begin{equation}\label{eq:vorthogonal}
\begin{array}{l}
\displaystyle\int_{V_{r_0}}\bm{v}_{\tau lm}^*(k{\bm{r}})\cdot\bm{v}_{\tau^\prime l^\prime m^\prime}(k{\bm{r}}){\rm d}v\vspace{0.2cm}\\
\hspace{2cm}=\displaystyle\delta_{\tau\tau^\prime}\delta_{ll^\prime}\delta_{mm^\prime}
\int_{V_{r_0}}|\bm{v}_{\tau lm}(k\bm{r})|^2{\rm d}v,
\end{array}
\end{equation}
where $\tau=1,2$.

The out-going (radiating) and in-going vector spherical waves $\bm{u}_{\tau lm}(k{\bm{r}})$ and $\bm{w}_{\tau lm}(k{\bm{r}})$ are obtained by replacing
the spherical Bessel functions ${\rm j}_l(x)$ above for the spherical Hankel functions of the first and second kind, ${\rm h}_l^{(1)}(x)$ and ${\rm h}_l^{(2)}(x)$, 
respectively, see \cite{Bostrom+Kristensson+Strom1991,Olver+etal2010}. The dagger notation $\{\cdot\}^\dagger$ is used here to denote a sign-shift in the exponent of the factor
$\eu^{\iu m\phi}$. Hence, for real arguments $kr$, it is observed that
\begin{equation}\label{eq:daggerrel}
\begin{array}{l}
\bm{v}_{\tau lm}^*(k\bm{r})=\bm{v}_{\tau lm}^\dagger(k\bm{r}), \vspace{0.2cm} \\
\bm{u}_{\tau lm}^*(k\bm{r})=\bm{w}_{\tau lm}^\dagger(k\bm{r}), \vspace{0.2cm} \\
\bm{w}_{\tau lm}^*(k\bm{r})=\bm{u}_{\tau lm}^\dagger(k\bm{r}),
\end{array}
\end{equation}
where $\{\cdot\}^*$ denotes the complex conjugate.

The free-space Green's dyadic for the electric field satisfies 
$\nabla\times\nabla\times \bm{G}_{\rm e}(k,\bm{r},\bm{r}^\prime)-k^2\bm{G}_{\rm e}(k,\bm{r},\bm{r}^\prime)={\bm I}\delta(\bm{r}-\bm{r}^\prime)$,
where $\bm{I}$ is the identity dyadic and $\delta(\cdot)$ the Dirac delta function, and is given by
\begin{equation}\label{eq:Greensdyadic1}
\bm{G}_{\rm e}(k,\bm{r},\bm{r}^\prime)=(\bm{I}+\frac{1}{k^2}\nabla\nabla)\frac{\eu^{\iu k|\bm{r}-\bm{r}^\prime|}}{4\pi|\bm{r}-\bm{r}^\prime|},
\end{equation}
see \eg \cite{Jackson1999,Bostrom+Kristensson+Strom1991}.
The free-space Green's dyadic can also be expanded in vector spherical waves as \eg
\begin{equation}\label{eq:Greensdyadic2}
\bm{G}_{\rm e}(k,\bm{r},\bm{r}^\prime)=\iu k\sum_{\tau=1}^2\sum_{l=1}^{\infty}\sum_{m=-l}^{l}\bm{u}_{\tau lm}(k\bm{r}_{>})\bm{v}_{\tau lm}^\dagger(k\bm{r}_{<}),
\end{equation}
where $\bm{r}_{>}$ ($\bm{r}_{<}$) denotes the vector in $\{\bm{r},\bm{r}^\prime\}$ having the largest (smallest) length, \cf \cite{Bostrom+Kristensson+Strom1991}.

\bibliographystyle{teorel}

\end{document}